\documentclass[useAMS,usenatbib]{mn2e}

\input{epsf}

\usepackage{graphicx}	
\usepackage{deluxetable}
\usepackage{amsmath}
\usepackage{amssymb}

\newcommand{\kms}{km~s$^{-1}$}

\newcommand{\OIIIw}{[O$\,\textsc{iii}]$~$\lambda$5007}

\newcommand{\Hb}{{H}\,$\beta$}
\newcommand{\FeII}{Fe\,{\sc ii}}

\newcommand{\NII}{[N\,{\sc ii}]}

\newcommand{\OIII}{[O\,{\sc iii}]}

\def\lsim{\lower0.3em\hbox{$\,\buildrel <\over\sim\,$}}
\def\gsim{\lower0.3em\hbox{$\,\buildrel >\over\sim\,$}}

\newcommand{\vmax}{v_\mathrm{max}}

\title[Rest-frame optical properties of RL BALs]{Rest-frame optical properties of luminous, radio-selected broad absorption line quasars}
\author[J. C. Runnoe et al.]{Jessie C. Runnoe$^{1}$\thanks{E-mail:
jrunnoe@uwyo.edu} , R. Ganguly$^{2}$, M. S. Brotherton$^{1}$, M.A. DiPompeo$^{1}$\\
$^{1}$Department of Physics and Astronomy, University of Wyoming, Laramie, WY 82071, USA\\
$^{2}$Department of Computer Science, Engineering, \& Physics, University of Michigan-Flint, Flint, MI 48503}

\begin{document}	

\date{Preprint 2012 December 11}

\pagerange{\pageref{firstpage}--\pageref{lastpage}} \pubyear{2012}

\maketitle

\label{firstpage}

\begin{abstract}
We have obtained IRTF/SpeX spectra of eight moderate-redshift ($z=0.7-2.4$), radio-selected (log$\,R^*\approx0.4-1.9$) broad absorption line (BAL) quasars. The spectra cover the rest-frame optical band. We compare the optical properties of these quasars to those of canonically radio-quiet (log$\,R^*\lesssim1$) BAL quasars at similar redshifts and to low-redshift quasars from the Palomar-Green catalog. As with previous studies of BAL quasars, we find that \OIIIw\ is weak, and optical \FeII\ emission is strong, a rare combination in canonically radio-loud (log$\,R^*\gtrsim1$) quasars. With our measurements of the optical properties, particularly the Balmer emission line widths and the continuum luminosity, we have used empirical scaling relations to estimate black hole masses and Eddington ratios. These lie in the range $(0.4-2.6) \times 10^9 M_\odot$\ and 0.1--0.9, respectively.  Despite their comparatively extreme radio properties relative to most BAL quasars, their optical properties are quite consistent with those of radio-quiet BAL quasars and dissimilar to those of radio-loud non-BAL quasars. While BAL quasars generally appear to have low values of \OIIIw/\FeII\, an extreme of ``Eigenvector 1'', the Balmer line widths and Eddington ratios do not appear to significantly differ from those of unabsorbed quasars at similar redshifts and luminosities.
\end{abstract}


\section{introduction}	
In the past decade or so, a new class of AGN -- radio-selected, broad absorption line (BAL) quasars -- has brought new insights into the structure of quasars and quasar outflows. The deep images of the Faint Images of the Radio Sky at Twenty-Centimeters survey \citep[FIRST;][]{becker95} and the NRAO VLA Sky Survey \citep[NVSS;][]{condon98} led to the detection of BAL quasars in sufficient numbers \citep[e.g.,][]{gregg96,becker97,brotherton98b,white00,becker00,becker01} that techniques for gauging orientation from radio properties (such as morphology, core-dominance, and spectral index) could be used test ideas about the structure of quasar outflows. 

\citet{becker00} reported that radio-selected BAL quasars had radio morphologies consistent with the entire range of viewing angles of Type 1 quasars (those exhibiting broad emission lines).  A comparison of the radio spectral index distributions between BAL and non-BAL quasars supports this finding \citep{becker00,montenegro08,fine11,dipompeo11a}.  Spectral index is a useful ensemble orientation indicator, as sources seen closer to the jet axis are dominated by radio core emission and will have flatter spectra than those seen more ``edge-on'' that are dominated by radio lobe emission.  Modeling spectral index distributions showed that while the likelihood of seeing a BAL increases at the largest inclinations, BAL outflows are found at a wide range of sight lines \citep{dipompeo12a}.  Studies of the optical/ultraviolet polarization have been used to argue for an edge-on orientation for BAL quasars, but when combined with radio information it is apparent that the situation is much more complex \citep{dipompeo10,dipompeo13a}.  Furthermore other results, such as the dependence of BAL fraction on redshift \citep{allen11}, suggest an evolutionary aspect to the BAL phenomenon.  \citet{shankar08} use the dependence of radio power on BAL strength to argue that no single model (orientation or evolution) is sufficient to explain the class.  All of these results continue to challenge the view that BAL quasars are simply ``normal'' Type 1 quasars seen at large viewing angles.

Since most, if not all, BAL quasars prior to FIRST and NVSS were radio-quiet \citep[e.g.,][]{stocke92}, it seems reasonable to ask the question whether these radio-selected BAL quasars are special in some way, different than their radio-quiet kin, and hence should be interpreted separately. We note here that, while the BAL quasars from FIRST are radio-{\it selected}, most are not radio-{\it loud} by canonical measures \citep[e.g., radio-to-optical flux ratios,][]{kellerman94}. Moreover, the higher levels of extinction of the optical flux typically found in BAL quasars and beaming of the radio flux may mean that some of these radio-selected BAL quasars are in reality intrinsically radio-quiet. Nevertheless, the question remains: are radio-selected BAL quasars more like other BAL quasars that happen to be detected in the radio, or are they more like radio-loud quasars that happen to have high velocity outflows? 

Clues to possible distinctions between radio-selected BAL quasars and radio-quiet BAL quasars have been looked for both in the X-ray \citep[e.g.,][]{brotherton05,miller09} and with spectropolarimetry \citep[e.g.,][]{brotherton97,brotherton06,dipompeo10}.  \citet{dipompeo12b} find that radio-selected BAL and non-BAL quasars have the same similarities and differences in rest-frame ultraviolet spectral properties as radio-quiet BAL versus non-BAL sources \citep[e.g.,][]{weymann91}. Here, we add more fuel to this discussion in presenting rest-frame optical spectra of a sample of radio-selected BAL quasars.

In presenting and discussing the optical properties of radio-selected BAL quasars from an observational viewpoint, we couch our analysis in the context of the \citet{bg92} Eigenvector 1.  \citet{bg92} used principal component analysis to parameterize a variety of optical spectral measurements of a complete sample of $z < 0.5$ quasars from the Palomar-Green survey into perpendicular eigenvectors that account for the largest amount of variance between quasar spectra.  The strongest set of relationships involves the strength of optical \FeII\ broad emission lines, the strength of narrow \OIIIw\ emission, and the width and asymmtery of the broad H$\beta$\ emission line. In short, the strongest variations in optical quasar spectra describe the change in objects with weak \FeII, strong \OIIIw, and broad H$\beta$\ to objects with strong \FeII, weak \OIIIw, and narrow  H$\beta$\ with strong blue wings. This set of relationships is collectively termed ``Eigenvector 1" (hereafter EV1) and is also related to properties in other wavebands.  Notably, radio-quiet quasars and BALs tend to sit at the end of EV1 with small EW(\OIIIw/\FeII), while the radio-loud quasars tend to sit at the other end with large EW(\OIIIw/\FeII).  The physical driver of EV1 remains elusive, although the Eddington ratio may play a role \citep{boroson02}.  In the \citet{bg92} analysis, Eigenvector 2, which is primarily driven by luminosity, and later Eigenvectors describe second order effects in quasar spectra and are less interesting in the context of RL BALs.

Observationally, the handful of radio-quiet broad-absorption line (BAL) quasars from the $z<0.5$\ Palomar-Green catalog all preferentially appear at the extreme radio-quiet end of EV1 (weak \OIII, strong \FeII, narrow H$\beta$).  At higher redshift, \citet{mcintosh99} finds the same behavior where radio-quiet and BAL objects sit on the opposite end of EV1 from radio-loud objects.  \citet{yuan03} (hereafter YW) investigated the properties of more luminous radio-quiet BAL quasars at $z\sim2$, finding that they also display extreme \OIII\ and optical \FeII\, but not extreme H$\beta$\ widths. Furthermore, \citet{ganguly07b} have shown that $z=1.7-2$\ BAL quasars are not extreme Eddington ratio objects, at least not in comparison to other luminous quasars at those redshifts. Both YW and \cite{ganguly07b} have proposed that, while Eddington ratio may be important to EV1 at low redshifts, other factors like absolute fuelling rate and environmental factors may play a role at higher redshifts and luminosities, and the relationships of EV1 may change.

Given the extremely different spectral characteristics of RL and BAL quasars, our primary goal is to determine whether radio-selected BAL quasars are more ``BAL-like'' or more ``RL-like'' in their optical properties.  We further investigate the implications of the behavior of radio-selected BAL optical properties for the bigger picture of BAL orientations and accretion physics, expressed in terms of EV1.

This makes RL BALs particularly interesting in terms of EV1, since they cannot simultaneously have large and small values of EW(\OIIIw/\FeII).

In \S\ref{sec:data}, we describe our sample, and present our IRTF data. In \S\ref{sec:analysis}, we detail the measurements made and physical quantities inferred from those measurements. In addition, we compute a composite optical spectrum of radio-selected BAL quasars. In \S\ref{sec:discussion}, we discuss the implications of our results both for quasar outflow geometry and for the interpretation of EV1, and we summarize our findings in \S\ref{sec:conclusion}. Throughout this paper, we adopt a cosmology with $H_0 = 71$\,\kms\,Mpc$^{-1}$, $\Omega_\Lambda = 0.73$, and $\Omega_m = 0.27$.

\section{Data}
\label{sec:data}
\subsection{Observations}

Our sample is a subset of the radio-selected BAL quasars from the FIRST Bright Quasar Survey from \citet{becker00}. The parent sample contains 29 quasars that are simultaneously blue, starlike sources and detected in the FIRST radio survey.  The observed subset of the parent sample was determined by taking high priority objects where H$\alpha$ and \Hb\ were well-placed in the available wave bands before bad weather prevented further data collection.  Some general information on these targets is listed in Table~\ref{tab:prop} including radio properties. BAL properties from \citet{becker00} are summarized in Table~\ref{tab:balprop}.  We observed eight of the brightest objects from this sample that also had redshifts putting optical emission lines into observable near-infrared windows. We made our observations over 27-28 Apr 2001 with the SpeX \citep{rayner03} instrument at the NASA Infrared Telescope Facility (IRTF) covering the wavelength range 0.8--2.4\,$\mu$m.  We employed a $0.\!\!''8$\ slit, and individual exposure times of 120s to avoid saturating the detector with background photons. Total exposure times are listed in Table~\ref{tab:obslog}. Nearly all observations were carried out with a position angle of 90$^\circ$. Q\,$1044+3656$\ was observed at a position angle of 155$^\circ$.

\begin{table*}
\begin{minipage}[2cm]{14.5cm}
\caption{Quasar general properties}
\label{tab:prop}
\renewcommand{\thefootnote}{\alph{footnote}}
\begin{tabular}{lcccccccccccc}

                     & 	&  \multicolumn{5}{c|}{Magnitudes}  				&   &  log L$_\nu$(5\,Ghz) 	&    			&            	&         A$_{v}$\footnotemark[3]         &     \\          
                            \cline{3-7}                              
Object         &	& $z$ & $J$ & $H$ & $K_\mathrm{s}$ & M$_{B}$ 	&  & (erg~s$^{-1}$~Hz$^{-1}$)	& log R$^{*}$\footnotemark[1] 		& $\alpha$\footnotemark[2] &    (mags) & Redshift  \\
\hline
0809+2753 && 17.34 & 16.25   & 15.65   & 15.55   & $-27.7$ 	&& 31.9 & 0.4  & \nodata &  0.00      &1.511  \\
1031+3953 && 18.01 & \nodata & \nodata & \nodata & $-25.7$&& 31.9 & 1.11 & $-0.2$  & 0.00      & 1.082  \\
1044+3656 && 16.59 & 15.55   & 15.16   & 14.34   & $-26.0$    && 32.1 & 1.29 & $-0.5$  & 0.21      & 0.701  \\
1312+2319 && 17.17 & 16.27   & 15.45   & 15.41   & $-27.4$ 	&& 33.3 & 1.88 & $-0.8$  &  0.00     & 1.508  \\
1324+2452 && 17.79 & 16.80   & 15.96   & 15.45   & $-27.5$ 	&& 32.7 & 1.31 & $-0.7$  & \nodata & 2.357  \\
1408+3054 && 17.21 & 16.10   & 15.65   & 15.00   & $-24.6$ 	&& 31.6 & 1.28 & $-0.7$  &  0.21      & 0.842  \\
1427+2709 && 17.96 & \nodata & \nodata & \nodata & $-25.5$&& 31.9 & 1.23 & $-0.7$  &  0.62      & 1.170  \\
1523+3914 && 16.32 & 15.36   & 14.88   & 13.87   & $-26.0$ 	&& 31.5 & 0.66 & $-0.4$  &  0.40      & 0.657  \\
\hline
\end{tabular}
\footnotetext[0]{Note $-$  Information for columns 6--10 was taken from Table~1 and 2 of \citet{becker00}, corrected for a $H_o = 71$\,\kms\,Mpc$^{-1}$, $\Omega_\Lambda = 0.73$, and $\Omega_m = 0.27$\ cosmology. }
\footnotetext[1]{The radio-loudness parameter, R*, is the ratio of the 5\,GHz radio flux density to the 2500\,\AA\ optical flux density in the quasar rest-frame.}
\footnotetext[2]{Spectral index ($F_\nu \sim \nu^\alpha$) in the radio band between 3.6\,cm and 20\,cm.}
\footnotetext[3]{$V$-band reddening calculated in \citet{dipompeo13b}.}
\end{minipage}
\end{table*}

\begin{table}
\begin{minipage}{6.5cm}
\renewcommand{\thefootnote}{\alph{footnote}}
\caption{BAL Properties \label{tab:balprop}}
\begin{tabular}{crrr}

Object & BALnicity & v$_\mathrm{max}$ & BAL Class      \\
\hline
0809+2753 & 7000 & 27400 & HiBAL   \\
1031+3953 & 20   & 5900  & LoBAL   \\
1044+3656 & 400  & 6600  & FeLoBAL \\
1312+2319 & 1400 & 25000 & HiBAL   \\
1324+2452 & 1300 & 6900  & LoBAL   \\
1408+3054 & 4800 & 22000 & LoBAL   \\
1427+2709 & 30   & 5900  & FeLoBAL \\
1523+3914 & 3700 & 19000 & LoBAL \\
\hline
\end{tabular}
\footnotetext[0]{Note $-$ All information is taken from Table~1 of \citet{becker00}.  v$_\mathrm{max}$ is in units of km s$^{-1}$.}
\end{minipage}
\end{table}
\begin{table*}
\begin{minipage}{11cm}
\renewcommand{\thefootnote}{\alph{footnote}}
\caption{IRTF/SpeX Observation Log \label{tab:obslog}}
\begin{tabular}{ccccc}
Object & R.A.          & Decl.         & Date & Exposure Time      \\
            & (J2000.0) & (J2000.0) &           & (s) \\
\hline
0809+2753 & 08 09 01.332 & +27 53 41.67 & 2001 April 28 & 5640 \\
1031+3953 & 10 31 10.647 & +39 53 22.81 & 2001 April 28 & 7200 \\
1044+3656 & 10 44 59.591 & +36 56 05.39 & 2001 April 27 & 3840 \\
1312+2319 & 13 12 13.560 & +23 19 58.51 & 2001 April 27 & 7440 \\
1324+2452 & 13 24 22.536 & +24 52 22.25 & 2001 April 28 & 7200 \\
1408+3054 & 14 08 06.207 & +30 54 48.67 & 2001 April 28 & 2400 \\
1427+2709 & 14 27 03.637 & +27 09 40.29 & 2001 April 27 & 6000 \\
1523+3914 & 15 23 50.435 & +39 14 04.83 & 2001 April 27 & 1200 \\
          	   &              	    &              	     & 2001 April 28 & 2400 \\

\hline
\end{tabular}
\footnotetext[0]{Note $-$ In all cases, exposure times were broken up into 120\,s segments to avoid background saturation.}
\end{minipage}
\end{table*}

We note that the BAL properties of the quasar $1408+3054$ are known to vary, with the UV iron absorption disappearing from the spectrum over a period of years \citep{hall11}.  Though it is listed as a low-ionization BAL (LoBAL) by \citet{becker00}, it is likely that it was in an iron low-ionization BAL (FeLoBAL) state at the time that our observations were made.  Spectra taken in 2000 and 2001, near the time of our observations, are presented in \citet{hall11} and \citet{dipompeo10}, respectively.  In both cases the UV \FeII\ absorption is significant.

\subsection{Data reduction}

We extracted the IRTF/SpeX spectra using the IDL-based Spextool software version 2.1 \citep{cushing04}. The apertures of different orders were located and traced using an internal template that is specific to the IRTF/SpeX instrument. The background is also fitted and subtracted. In cases where the target was too faint and the Spextool algorithm had difficulty in finding and tracing the apertures, we used the fitted aperture trace function from a close-by atmospheric star as a template to extract the spectral data from the target frame.

Spextool also performs wavelength calibration. The target frame extraction parameters were used as templates to extract spectra from the calibration lamp frames. We reviewed the line identification on the calibration frames, fit a dispersion curve, and then applied the wavelength solution to the
target spectra.

For relative flux calibration within orders and in between orders, we assumed that both the quasar spectra and the F or G standard star spectra were affected by the same atmospheric absorption and detector response functions. We divided the observed quasar spectra by appropriate standard star spectra to remove the effect of the telluric absorption and response functions \citep{vacca03}. Then we multiplied the result with the assumed black body spectrum of the standard star to get the relative flux density spectrum of the target quasar. Possible source of uncertainties in this step are as follows.

\begin{itemize}

\item The removal of stellar absorption lines from the standard star, especially in the atmosphere absorbed parts of the spectrum, might be incomplete. Since the absorption lines in F and G stars are very weak and their peak intensity is less than 5\% of the continuum level in most cases, this only produces $<$5\% spikes in the noisy regions of the final spectrum.
\item The F and G standard star intrinsic continuum shapes are not perfect blackbodies.  This introduces about a 2$-$3\% of error in the continuum shape.
\end{itemize}

For each IRTF/SpeX object, we combined spectra of different orders to form a 1-D continuous spectrum. We verified that the regions with overlapping wavelengths agreed with each other within 3$\sigma$ of the noise. In those overlapping spectral regions, we calculated average values of the flux for the combined spectrum. As expected, the combined spectra showed very high noise levels outside the traditional atmospheric windows.

To carry out absolute flux calibration, we matched the observed count rate in the band containing the H$\alpha$\ emission line (where we usually have the most signal) to the corresponding J, H, or K$_\mathrm{s}$\ 2MASS magnitudes \citep{skrutskie06}. In two cases, Q\,$1031+3953$\ and Q\,$1427+2709$, we used the Sloan z magnitude \citep[e.g., ][]{schneider07,fukugita96} as the quasars were not detected by 2MASS. There was sufficient overlap between the Sloan z bandbass and the blue side of the SpeX spectrum to do so.

\begin{figure*}
\begin{center}
\includegraphics[width=18 truecm]{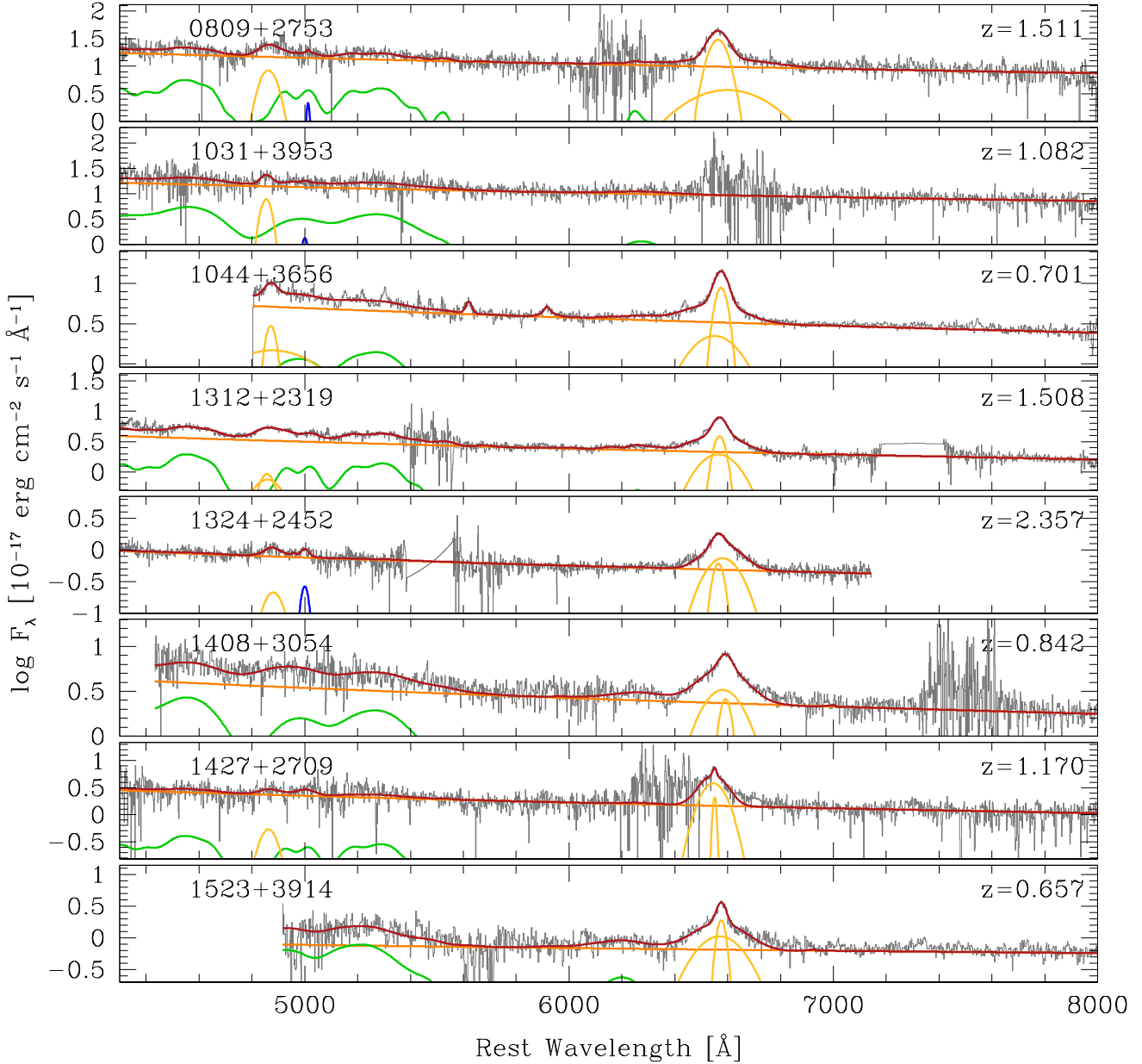}
\end{center}
\caption{Rest-frame optical spectra of objects in our sample. IRTF/SpeX spectra have been flux-calibrated to either 2MASS or SDSS photometry.  Data are shown as a gray histogram. Superimposed on on the data is the total fit from our {\sc specfit} model (red), and the individual components of that model (orange: power-law continuum; green: \FeII\ template; yellow: Balmer emission-line components; blue: \OIII\ emission-line components) which are given arbitrary normalizations to help visualize the fit..}
\label{fig:specfit}
\end{figure*}

The IRTF/SpeX spectra are shown in Figure~\ref{fig:specfit}.  For rest-frame UV spectra for all but one of these objects, see \citet{dipompeo10} or \citet{becker00,becker01}.

\section{Analysis}
\label{sec:analysis}

\subsection{Spectral fitting and derived parameters}

Our primary goal is to make measurements of the rest-frame optical continuum and emission-lines present in the SpeX spectra. We use the IRAF\footnote{IRAF is distributed by the National Optical Astronomy Observatories, which are operated by the Association of Universities for Research in Astronomy, Inc., under cooperative agreement with the National Science Foundation.} package {\sc specfit} \citep{kriss94} to carry out these measurements. We model the optical spectrum as a power-law continuum, with superimposed Gaussians for the \OIII\ $\lambda$4959, 5007, H$\alpha$, and H$\beta$\ emission lines, and the \FeII\ emission-line template from I Zw 1 \citep{bg92}. We do not attribute physical meaning to the Gaussians, but merely use them to reproduce and characterize the emission lines. For the \OIII\ emission lines, we use two Gaussians, one for each component of the doublet; we use two Gaussians (both generally broad with $>2500$~km~s$^{-1}$) to reproduce the H$\alpha$\ and H$\beta$\ emission lines.  All objects in this sample have extremely weak contributions from the narrow line region, so we do not include such a component in the \Hb\ fit.  Similarly, the \NII\ emission that often flanks the H$\alpha$\ emission in AGN is not readily seen in our spectra and we do not include it in our fitting model.  Our fits are shown in Fig.~\ref{fig:specfit}, and the measurements are listed in Tables~\ref{tab:specfit1} and \ref{tab:specfit2}.  When more than one Gaussian was used for a line fit, the parameters listed in these tables are for the combined line profile.  Note that the \Hb\ emission in these objects is relatively weak and in most cases, the \Hb\ fits are more uncertain than the formal errors indicate.  We include them here for completeness and because they are not independent of the other fit components, but we generally use H$\alpha$ for making physical calculations.

We have not included intrinsic reddening in our fits.  The objects in this sample are all UV-selected, implying that they cannot have significant reddening at UV wavelengths.  This is confirmed by the estimates of reddening available from \citet{dipompeo13b} for all but one of the objects that are listed in Table~\ref{tab:prop}.  Briefly, these values are estimated by determining by eye the amount of reddening required to match each object's spectrum to the composite spectrum of all FIRST Bright Quasar Survey (FBQS) quasars \citep{white00, brotherton01} using a Small Magellanic Cloud extinction curve.  A more detailed description is available in \citet{dipompeo13b}.  The majority of our sample demonstrates insignificant reddening in the $V$ band, and in all cases the optical reddening at 5100 \AA\ should be negligible.

Formal errors on the fit parameters are calculated following the method of \citet{dipompeo12b}.  We created synthetic spectra of each object by calculating the noise in each spectrum between 5090 and 5110 \AA\ and then added random Gaussian noise consistent with this estimate to the best-fitting model.  We then fit the synthetic spectra and calculated spectral parameters following the procedure described above.  This process is repeated 50 times per object and the standard deviation of each fit parameter is taken as the uncertainty.  We stress that these errors are the formal uncertainties obtained from this fitting procedure but following another fitting procedure may yield results that differ from ours by more than these errors.  In fact, particularly for \OIII\ and \Hb\ emission which is often weak and located in a spectral region with greater noise, the formal errors likely underestimate the true uncertainty.

\begin{table*}
\begin{minipage}{17.5cm}
\caption{Spectral parameters of optical \FeII\ blend and \OIII}
\renewcommand{\thefootnote}{\alph{footnote}}
\label{tab:specfit1}
\begin{tabular}{lcccccccccccccccccccccc}
&&&  \multicolumn{2}{c|}{Power Law}  &&  \multicolumn{2}{c|}{Optical \FeII} && \multicolumn{3}{c|}{\OIII\ $\lambda 5007$, $\lambda 4959$\footnotemark[1]} \\
\cline{4-5}\cline{7-8}	 \cline{10-12}
Object  & Redshift && $f_{1000}$ & $\alpha$ && Scale 	& FWHM && Flux & FWHM & Centroid  \\
\hline
0809+2753
	& 
   1.511
 && 
   12.70
$\pm$    0.02
 & 
    1.36
$\pm$    0.01
 && 
   0.258
$\pm$   0.023
 & 
        3478
$\pm$         780
 && 
    3.05
$\pm$    0.09
 & 
         791
$\pm$           7
 & 
        5013
$\pm$    0.02
 \\
1031+3953
	& 
   1.082
 && 
   11.70
$\pm$    0.29
 & 
    1.35
$\pm$    0.01
 && 
   0.288
$\pm$   0.016
 & 
        9000

 && 
    4.09
$\pm$    0.09
 & 
        1730
$\pm$           1
 & 
        5000
$\pm$    0.06
 \\
1044+3656
	& 
   0.701
 && 
    5.63
$\pm$    0.04
 & 
    1.51
$\pm$    0.01
 && 
   0.094
$\pm$   0.003
 & 
        2672
$\pm$          18
 && 
\nodata

 & 
\nodata

 & 
\nodata

 \\
1312+2319
	& 
   1.508
 && 
    3.18
$\pm$    0.12
 & 
    1.43
$\pm$    0.02
 && 
   0.091
$\pm$   0.003
 & 
        3958
$\pm$         114
 && 
    0.04
$\pm$    0.06
 & 
        1029
$\pm$          10
 & 
        5010
$\pm$    0.14
 \\
1324+2452
	& 
   2.357
 && 
    0.99
$\pm$    0.04
 & 
    1.60
$\pm$    0.02
 && 
   0.002
$\pm$   0.001
 & 
        2777
$\pm$          47
 && 
    0.89
$\pm$    0.12
 & 
        2025
$\pm$         317
 & 
        5000

 \\
1408+3054
	& 
   0.842
 && 
    3.74
$\pm$    0.26
 & 
    1.46
$\pm$    0.04
 && 
   0.128
$\pm$   0.012
 & 
        2505
$\pm$         117
 && 
\nodata

 & 
\nodata

 & 
\nodata

 \\
1427+2709
	& 
   1.170
 && 
    2.85
$\pm$    0.05
 & 
    1.58
$\pm$    0.02
 && 
   0.017
$\pm$   0.005
 & 
        1000

 && 
    3.42
$\pm$    2.79
 & 
        3957
$\pm$         779
 & 
        5007
$\pm$    0.93
 \\
1523+3914
	& 
   0.657
 && 
    0.21
$\pm$    0.01
 & 
    0.63
$\pm$    0.01
 && 
   0.056
$\pm$   0.004
 & 
        7678
$\pm$           2
 && 
\nodata

 & 
\nodata

 & 
\nodata

 \\
Composite
	& 
\nodata
 && 
   17.90
$\pm$    0.45
 & 
    1.63
$\pm$    0.01
 && 
   0.233
$\pm$   0.008
 & 
        2474
$\pm$           1
 && 
    0.65
$\pm$    0.85
 & 
         852
$\pm$          22
 & 
        5006
$\pm$    0.09
 \\
FBQS Composite
	& 
\nodata
 && 
    0.78
$\pm$    0.01
 & 
    0.12
$\pm$    0.01
 && 
   0.069
$\pm$   0.001
 & 
        3786
$\pm$          57
 && 
   32.83
$\pm$    0.06
 & 
         739
$\pm$           1
 & 
        5006
$\pm$    0.01
 \\
PG Composite
	& 
\nodata
 && 
   15.38
$\pm$    0.13
 & 
    1.64
$\pm$    0.00
 && 
   0.194
$\pm$   0.001
 & 
        3602
$\pm$          57
 && 
    4.18
$\pm$    0.08
 & 
         811
$\pm$          19
 & 
        5006
$\pm$    0.15
 \\
\hline
\end{tabular}
\footnotetext[0]{Note $-$ The power law is given as $f_{1000} \sim \lambda^{-\alpha}$, with flux density units $10^{-16}$ erg cm$^{-2}$ s$^{-1}$ \AA$^{-1}$. For the emission-line components, we list the integrated flux (in $10^{-16}$\,erg~cm$^{-2}$~s$^{-1}$), the centroid (in \AA), and the full-width at half maximum (in km s$^{-1}$) of the combined line profile. For the \FeII\ template, the flux listed is relative to the strength of the I\,Zw\,1 \FeII\ emission. When no values are reported for a given component, that component was not used in the fit. When no errors are reported for a given component, that component reached a limit and became fixed in the fit.  We report the redshift as determined by the centroid of the total H\,$\alpha$\ line. In the case of Q\,$1031+3953$, where the H\,$\alpha$\ is not covered, we use the \OIIIw\ line. All wavelengths are reported relative to our optically-based redshift. The optically-based redshift is also employed in the construction of the composite.}
\footnotetext[1]{The \OIII\ doublet was fit with two Gaussian components.  The $\lambda 4959$ component has the listed flux and FWHM with a centroid of 0.9904272 times the $\lambda 5007$ centroid.}
\end{minipage}
\end{table*}

\begin{table*}
\begin{minipage}{17.5cm}
\caption{Spectral parameters of H$\alpha$ and H$\beta$}
\renewcommand{\thefootnote}{\alph{footnote}}
\label{tab:specfit2}
\begin{tabular}{lcccccccccccccccccccccc}
&&&  \multicolumn{3}{c|}{H$\beta$ \footnotemark[1]}  	&&  \multicolumn{3}{c|}{H$\alpha$}  \\
\cline{4-6}\cline{8-10}
Object  & Redshift && Flux & FWHM & Centroid &&  Flux & FWHM & Centroid\\
\hline
0809+2753
	& 
   1.511
 && 
   71.92
$\pm$    0.15
 & 
        4780
$\pm$           1
 & 
        4862
$\pm$    2.14
 && 
  398.69
$\pm$    0.01
 & 
        3989
$\pm$           1
 & 
        6563
$\pm$    0.35
 \\
1031+3953
	& 
   1.082
 && 
   46.83
$\pm$    0.09
 & 
        2913
$\pm$           1
 & 
        4853
$\pm$    0.09
 && 
\nodata

 & 
\nodata

 & 
\nodata

 \\
1044+3656
	& 
   0.701
 && 
   82.46
$\pm$    0.64
 & 
        4381
$\pm$          43
 & 
        4871
$\pm$    0.66
 && 
  110.49
$\pm$    0.54
 & 
        3115
$\pm$           7
 & 
        6574
$\pm$    0.33
 \\
1312+2319
	& 
   1.508
 && 
   18.76
$\pm$    1.23
 & 
        5973
$\pm$         186
 & 
        4857
$\pm$    1.60
 && 
   61.83
$\pm$    0.36
 & 
        3362
$\pm$           9
 & 
        6569
$\pm$    0.18
 \\
1324+2452
	& 
   2.357
 && 
    2.50
$\pm$    0.15
 & 
        3619
$\pm$         473
 & 
        4873
$\pm$    1.54
 && 
   15.19
$\pm$    0.10
 & 
        4044
$\pm$          37
 & 
        6566
$\pm$    0.07
 \\
1408+3054
	& 
   0.842
 && 
   16.52
$\pm$    8.59
 & 
        7731
$\pm$        5510
 & 
        4867
$\pm$   14.08
 && 
   80.86
$\pm$    0.48
 & 
        4650
$\pm$          36
 & 
        6590
$\pm$    0.14
 \\
1427+2709
	& 
   1.170
 && 
    4.29
$\pm$    2.08
 & 
        4977
$\pm$           1
 & 
        4863
$\pm$    6.97
 && 
   48.05
$\pm$    2.57
 & 
        3063
$\pm$         729
 & 
        6550
$\pm$    2.82
 \\
1523+3914
	& 
   0.657
 && 
\nodata

 & 
\nodata

 & 
\nodata

 && 
   30.61
$\pm$    0.56
 & 
        2533
$\pm$          31
 & 
        6575
$\pm$    0.39
 \\
Composite
	& 
\nodata
 && 
   47.27
$\pm$    1.03
 & 
        4803
$\pm$          55
 & 
        4860
$\pm$    1.22
 && 
  280.05
$\pm$    1.71
 & 
        3672
$\pm$          25
 & 
        6566
$\pm$    0.73
 \\
FBQS Composite
	& 
\nodata
 && 
   65.99
$\pm$    0.67
 & 
        6819
$\pm$         458
 & 
        4864
$\pm$    0.58
 && 
  230.71
$\pm$    0.28
 & 
        4927
$\pm$          10
 & 
        6564
$\pm$    0.04
 \\
PG Composite
	& 
\nodata
 && 
   59.80
$\pm$    0.36
 & 
        3068
$\pm$          30
 & 
        4861
$\pm$    0.15
 && 
\nodata

 & 
\nodata

 & 
\nodata

 \\
\hline
\end{tabular}
\footnotetext[0]{Note $-$ Measurements are reported in the same way as in Table~\ref{tab:specfit1}.}
\footnotetext[1]{Fits to \Hb\ are more uncertain than the formal errors in most cases, but we report them because they are not independent of the other spectral components.  These values are not used for calculating physical properties except in the case of 1031+3953 where H$\alpha$ is absent and the \Hb\ fit is acceptable.}
\end{minipage}
\end{table*}

All of the spectra have well-detected H$\alpha$ except one, Q$1031+3953$, where the H$\alpha$ emission is buried in the atmospheric absorption.  The \Hb\ emission line is detected in all of the spectra except for Q$1523+3914$, where the coverage cuts off at wavelengths longer than \Hb.  The spectra are often noisy in the \Hb\ wavelength regime and the \Hb\ emission is much more difficult to resolve than the stronger H$\alpha$ emission.  The formal errors on the fit parameters for \Hb\ are generally small but measurements made via other fitting procedures may yield results inconsistent with ours even with the errors.  Thus, we rather calculate physical parameters from the H$\alpha$ emission line whenever possible.

Given our choice to compute physical parameters from the H$\alpha$ emission line, the calculation of physical parameters for this sample becomes somewhat non-standard so we describe it in detail below.  In order to compare to other samples we employ the H$\alpha$ - H$\beta$\ FWHM relation from \citet{shen11}:

\begin{eqnarray}
\label{eqn:FWHM}
\nonumber \rm{log}\left ( {{\rm{FWHM}(\rm{H}\beta)} \over {10^3 \mathrm{km~s}^{-1}}}
\right ) = (-0.11\pm0.03)+(1.05\pm0.01)\\
\times \, \rm{log}\left ( {{\rm{FWHM}(\rm{H}\alpha)} \over {10^3 \mathrm{km~s}^{-1}}}
\right ).
\end{eqnarray}

The H$\alpha$\ emission line FWHM is estimated from the sum of the two Gaussian components. The resulting H$\beta$\ FWHM are listed in Table~\ref{tab:phys} (column 2).  We draw attention to the fact that the values of FWHM for \Hb\ measured from the spectra are systematically larger than the calculated ones.  In the presence of strong \FeII\ emission, as is common in BALs, and weak \Hb\ emission in low signal-to-noise spectra, the fitting method that we employ will often create very broad \Hb\ in order to bury the \Hb\ emission in the \FeII\ and the noise.  We take the broad measured values of \Hb\ FWHM as evidence that this is occurring in our fitting, at least to some extent, and confirm out choice to measure physical parameters from the H$\alpha$ emission line.

\begin{figure}
\begin{center}
\includegraphics[width=8 truecm]{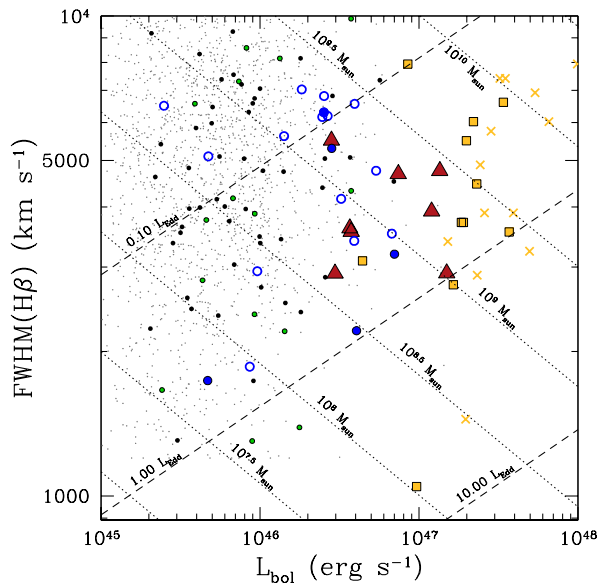}
\end{center}
\caption{We show the H$\beta$\ emission-line FWHM versus the bolometric luminosity for several samples of quasars (tiny gray points: radio-loud \citet{shen11}; larger black dots: \citet{shen11} BALs; green dots: radio-loud \citet{shen11} BALs; blue filled circles: BQS BALs; blue open circles: radio-loud BQS; yellow filled squares squares: YW BALs; yellow crosses: radio-loud YW; filled red triangles: radio-selected BALs from this work). Diagonal lines indicates black hole mass and Eddington ratio isopleths based on the \citet{vestergaard06} scaling relation. The bolometric luminosity is computed from the \citet{runnoe12a} bolometric correction to the 5100\,\AA\ luminosity.  The properties of radio-selected BALs do not stand out from the BQS and YW samples, and they are not super-accretors.}
\label{fig:hbfwhml}
\end{figure}

For black hole mass estimates and also Eddington ratios, there are robust mass scaling relationships based on the H$\alpha$ line.  \citet{greene05} and \citet{greene10a} provide scaling relations for the FWHM of H$\alpha$ and L$_{\textrm{H}\alpha}$ and $\lambda$L$_{\lambda}$ (5100 \AA), respectively.  All mass scaling relationships are the result of reverberation mapping, for which \Hb\ has the most results.  Thus, the H$\alpha$ scaling relationships are built by deriving a conversion between the FWHM of \Hb\ and the FWHM of H$\alpha$, similar to Equation~\ref{eqn:FWHM}.  When L$_{\textrm{H}\alpha}$ is used, it is calculated from $\lambda$L$_{\lambda}$ (5100 \AA).  We have chosen to use the more recent conversion from \citet{shen11} for calculating FWHM and avoid a second conversion to get the H$\alpha$ luminosity.  We are also motivated to use an \Hb\ black hole mass scaling relationship by Figure~\ref{fig:hbfwhml}; we present that figure in terms of \Hb\ because it is a more common measurement in the literature, so we are therefore constrained to use an \Hb\ black hole mass scaling relationship in order to be consistent with the lines of constant black hole mass drawn in the figure.  With \Hb\ parameters thus in hand, we use the \citet{vestergaard06} relation using the luminosity at 5100\,\AA\ and the H$\beta$\ FWHM calculated from Equation~\ref{eqn:FWHM}:

\begin{eqnarray}
\label{eqn:Mbh}
\nonumber \rm{log}(M_\mathrm{BH}) = \rm{log}\left \{ \left [\frac{\rm{FWHM}(\rm{H}\beta)}{1000\textrm{ km s}^{-1}}\right ]^2 \left [ \frac{\lambda\,L_{\lambda}(5100 \textrm{\AA})}{10^{44}\textrm{ ergs s}^{-1}}\right]^{0.5}\right \} \\
+\,(6.91\pm0.02).
\end{eqnarray}

We calculate the correction to the scaling relationship for cosmology and find that it is negligible.  Our black hole mass estimates are listed in Table~\ref{tab:phys} (column 3). For Eddington luminosity estimates, we use the relation $L_\mathrm{Edd} = 1.51 \times 10^{38} (M/M_\odot)$\,erg~s$^{-1}$\ \citep{krolik99}.

We estimate the bolometric luminosity of our objects by correcting the continuum luminosity at rest-frame 5100\,\AA\ following \citet{runnoe12a}.
\begin{eqnarray}
\label{eqn:lbol}
\rm{log}(L_{bol})=(4.891\pm1.657)+(0.912\pm0.037) \,\rm{log}(\lambda L_{\lambda})
\end{eqnarray}

These bolometric luminosity estimates as well as resulting estimating of the Eddington ratio are listed in Table~\ref{tab:phys} (columns 4 and 5).

\begin{table*}
\begin{minipage}{17.5cm}
\caption{Calculated BAL QSO properties}
\renewcommand{\thefootnote}{\alph{footnote}}
\label{tab:phys}
\begin{tabular}{lccccccc}
Object  & FWHM(H$\beta)_{calc}$\footnotemark[1] & log$(M_{BH})$ & log($L_{\rm{bol}}$) & $L_{\rm{bol}}/L_{\rm{edd}}$ & EW(\OIIIw)\footnotemark[2] & EW(\FeII)\footnotemark[2] & EW(\OIII/\FeII)\\
        & km s$^{-1}$	   & $M_{\odot}$& ergs s$^{-1}$ 	 & 			       & \AA\        & \AA\      &\\
\hline
0809+2753
	& 
        4688
 & 
    9.27
 & 
   46.87
 & 
   0.266
 & 
0.77$\pm$0.07
 & 
10.98$\pm$0.08
 & 
0.07
 \\
1031+3953
	& 
        2913
 & 
    9.02
 & 
   47.17
 & 
   0.944
 & 
1.15$\pm$0.09
 & 
92.0$\pm$1.0
 & 
0.01
 \\
1044+3656
	& 
        3615
 & 
    8.87
 & 
   46.57
 & 
   0.325
 & 
$<$2.0
 & 
44.90$\pm$0.10
 & 
$<$0.05
 \\
1312+2319
	& 
        3916
 & 
    9.23
 & 
   47.08
 & 
   0.473
 & 
$<$0.64
 & 
143.90$\pm$0.30
 & 
$<$0.04
 \\
1324+2452
	& 
        4755
 & 
    9.42
 & 
   47.13
 & 
   0.339
 & 
3.3$\pm$0.3
 & 
11.80$\pm$0.20
 & 
0.28
 \\
1408+3054
	& 
        5507
 & 
    9.18
 & 
   46.45
 & 
   0.124
 & 
$<$5.0
 & 
256.0$\pm$2.0
 & 
$<$0.02
 \\
1427+2709
	& 
        3552
 & 
    8.86
 & 
   46.57
 & 
   0.340
 & 
7.0$\pm$1.0
 & 
42.8$\pm$0.8
 & 
0.17
 \\
1523+3914
	& 
        2909
 & 
    8.63
 & 
   46.47
 & 
   0.456
 & 
$<$24.0
 & 
61.64$\pm$0.05
 & 
$<$0.38
 \\
\hline
\end{tabular}
\footnotetext[1]{This value of FWHM(\Hb) is calculated from FWHM(H$\alpha$) via the prescription of \citet{shen11}.  This is the value used to calculate black hole mass.}
\footnotetext[2]{Quoted equivalent widths are the integrated line flux (see text for \FeII) divided by the continuum flux at 4861$\,$\AA, as given by the power-law fit in Table~5.}
\end{minipage}
\end{table*}

We compare our sample to the higher-redshift sample from YW and the lower-redshift Bright Quasar Survey sample \citep[BQS,][]{bg92} and {\it Sloan Digital Sky Survey} (SDSS) quasars from \citet{shen11}.  In order to facilitate the comparison we re-calculate properties of interest for these samples in a consistent way.  To calculate monochromatic luminosity at 5100 \AA\ for the YW sample, we use the measured 3000\,\AA\ luminosities and a spectral index of $\alpha_\nu = -0.44$\ \cite[as is typical for quasars in this reshift range;][]{vandenberk01}.  For the BQS sample, we use the measured 9480 \AA\ luminosities and spectral indices from \citet{neugebauer87}.  \citet{shen11} provide a measure of $\lambda$L$_{\lambda}$(5100 \AA) that we adopt, although their average correction for host galaxy contamination warrants some discussion.  All of the objects that we include from \citet{shen11} have redshifts less than 0.89, where objects with redshifts less than 0.5 are at risk for significant host contamination.  In fact, the source luminosity is the important parameter for determining the fractional contribution to emission from the host galaxy, with low-luminosity sources prone to significant contamination.  \citet{shen11} show that for sources with log$\,\lambda$L$_{\lambda}$(5100 \AA)$<$44.5 the host contributes on the order of a few percent or less of the emission.  For our bolometric correction this corresponds to log$(L_{bol})\lesssim45.5$.  The average host correction employed by \citet{shen11} may not be particularly accurate for individual objects, so the low-luminosity SDSS points where the host contributed a significant fraction of the total emission may be much more uncertain.  Finally, to calculate bolometric luminosity in all samples we applied the bolometric correction in Equation~\ref{eqn:lbol}.  

The fact that our objects are not special in any particular parameter space is illustrated in Figure~\ref{fig:hbfwhml}, where we plot the H$\beta$\ FWHM versus bolometric luminosity (updating YW Figure~2). We have placed isopleths of black hole mass \citep[using the relation of][]{vestergaard06} and Eddington ratio in the figure.  The bolometric luminosities of our objects lie (mostly) below those of the higher-redshift YW sample and above the lower-redshift BQS and \citet{shen11} samples, as is completely expected for simple flux-limited surveys.  This figure also clearly shows that neither radio-selected BAL quasars nor radio-quiet BAL quasars are exclusively super-accretors with $L_\mathrm{bol}/L_\mathrm{Edd} \gtrsim 1$. We discuss this further below (\S\ref{sec:discussion}).

\begin{figure}
\begin{center}
\includegraphics[width=8 truecm]{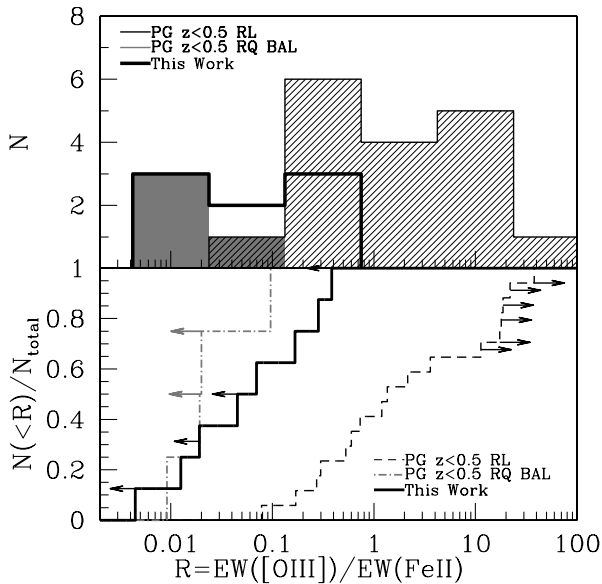}
\end{center}
\caption{We show the cumulative probability distribution of the EW(\OIII)/EW(\FeII) equivalent-width ratio for three samples: our sample of radio-selected BAL quasars (bold solid line); 17 low-redshift radio-loud objects from the Bright Quasar Survey \citep[BQS,][dashed black line]{bg92}; and 4 low-redshift, radio-quiet BAL quasars from the BQS (dot dashed gray line).  Arrows indicate objects that have only limits on the ratio from non-detection of either \OIII\ or \FeII. The limits are quoted at 3$\sigma$\ confidence.}
\label{fig:o3fe2rat}
\end{figure}

We also consider the equivalent width ratio between the \OIIIw\ emission-line and the optical \FeII\ emission-line complex. We follow the same method as \citet{bg92} to calculate equivalent widths, taking the integrated flux of the emission line divided by the continuum flux at 4861\,\AA\ (the middle of the H$\beta$\ emission line). For the \FeII\ flux, we integrate over the wavelength range 4434--4684\,\AA. In Figure~\ref{fig:o3fe2rat}, we plot the cumulative distributions of the ratio for our sample and two subsamples from BQS (radio-loud objects and radio-quiet BALs).  Note that the BQS radio-loud objects are typically more radio-loud and cover a larger range in log($R^{*}$) than our sample.  It is clear that the distribution of ratios for our radio-selected BALs differ greatly from the radio-loud objects in the BQS (with mean ratios of $\langle$EW(\OIII)/EW(\FeII)$\rangle\approx 0.1$\ and 1, respectively, and with only a 0.1\% KS probability of being drawn from the same distribution). On the other hand, the ratio resembles more the radio-quiet BALs from the BQS ($\langle$EW(\OIII)/EW(\FeII)$\rangle\approx 0.03$), with a KS probability of 74\% of being drawn from the same parent distribution.  The equivalent width ratio of \OIII\ to \FeII\ is essentially an estimate of an objects location on EV1, so this result supports the conclusion that RL BALs are more ``BAL-like'' than ``RL-like''.

\subsection{A composite optical spectrum}

To compare visually the general optical properties of radio-selected BAL quasars to other samples, we compute a composite spectrum from our eight quasars. To do so, we take the following steps: (1) Mask out pixels in the observed wavelength ranges 1.51--1.59\,$\mu$m and 2.03--2.16\,$\mu$m since these ranges cover the gaps between the J, H, and K bands. (2) Use a cubic spline interpolation to place all spectra on the same rest-frame wavelength scale. The scale covers the range 3500-9000\,\AA\ in 3\,\AA\ bins. (3) Normalize the 5900-6000\,\AA\ average flux to unity. (4) For each wavelength bin, median combine the normalized flux from all spectra that have not been masked out at that wavelength.

\begin{figure*}
\begin{center}
\includegraphics[width=15 truecm]{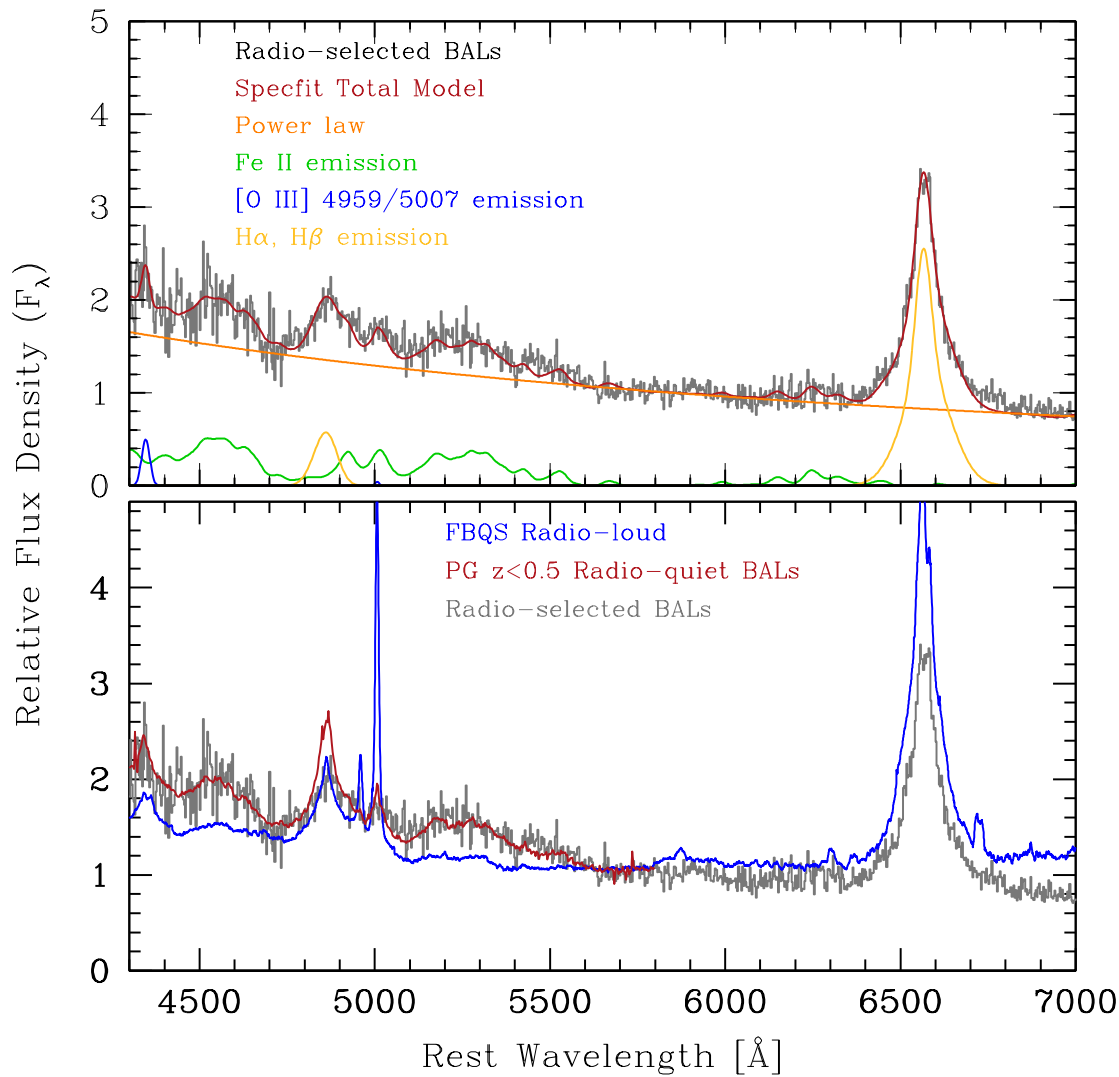}
\end{center}
\caption{We show a composite optical spectrum of the eight IRTF/SpeX spectra of radio-selected BAL quasars (gray, solid histogram). In the top panel of the figure, we overlay the results of our spectral fit [total fit (red), Balmer emission (yellow), \FeII\ template emission (green), \OIII\ emission (blue), power law continuum (orange)]. In the bottom panel, we show for comparison the composite quasar spectra of radio-loud quasars (blue) from the FIRST Bright Quasar Survey \citep{brotherton01}, as well as a composite of low-redshift, radio-quiet BAL quasars from the Bright Quasar Survey \citep[red][]{bg92}. All composites have been scaled to unit flux over the 5600--5680\,\AA\ wavelength range.}
\label{fig:composite}
\end{figure*}

The result of the compositing procedure is shown in Figure~\ref{fig:composite}. Two other composite spectra are also shown: radio-loud quasars from the FBQS \citep{brotherton01}, and a composite of BAL quasars from \citet{bg92} that have observed outflow velocities larger than 10,000\,\kms \citep[as listed in ][]{laor02}. For this latter composite, we included the following quasars: PG\,$1700+518$\ ($\vmax = 31,000$\,\kms), PG\,$2112+059$\ ($\vmax = 24,000$\,\kms), PG\,$0043+039$\ ($\vmax = 19,000$\,\kms), PG\,$1004+130$\ ($\vmax = 12,000$\,\kms), PG\,$1001+054$\ ($\vmax = 10,000$\,\kms). To compute this composite, we use the same procedure as above except for step three, where we used the 4500--4510\,\AA\ flux to normalize the spectra. In addition, we have made measurements of the optical properties of the all three composites using basically the same {\sc specfit} procedure as described above and listed them in Tables~\ref{tab:specfit1} and \ref{tab:specfit2} (the results of the fit are also shown in Figure~\ref{fig:composite}.  

There were some differences in how the FBQS composite was fit that resulted from the strong \OIII\ and weak \FeII.  In addition to the two generally broad Gaussians, the fits to the \Hb\ and H$\alpha$ emission lines included a narrow Gaussian to account for emission from the narrow-line region.  This contribution is not included in the presented fit measurements.  Additionally, the very strong $\lambda$5007 line required two Gaussians to achieve a good fit.  These are combined in the presented fit measurements.  In general for both the FBQS and PG composites, the \FeII\ template was not an excellent fit so the errors may be underestimated as a result.

\section{Results and discussion}
\label{sec:discussion}
\subsection{Results}
From our analysis of rest-frame optical spectra of radio-selected BAL quasars, our main results are:
\begin{itemize}
\item[1.] \OIII\ emission is weak, while optical \FeII\ emission is strong. The optical spectra of radio-{\it selected} BAL quasars are more similar to radio-{\it quiet} BAL quasars than RL quasars ($\langle$EW(\OIII])/EW(\FeII)$\rangle \approx 0.1$\ and $0.03$, respectively).

\item[2.] The physical properties of these quasars lie in the ranges: $0.1-0.9L_\mathrm{Edd}$, $(0.4-2.6) \times 10^9 M_\odot$.  We find that radio-selected BALs are not predominantly super-accretors.
\end{itemize}

There are two important comparisons to make in order to understand radio-selected BAL quasars and their role in the greater scheme of things. How do radio-selected BAL quasars compare to other BAL quasars? How do radio-selected BAL quasars compare to other radio-loud objects? Of particular interest is the finding from previous studies \citep[e.g.,][]{becker00} that radio-selected BAL quasars appear to show a range of orientations. What does this mean in terms of accretion physics? That is, do our objects add any insight into the interpretation of the \citet{bg92} EV1? In the following subsections, we discuss these two points in light of our results.

\subsection{The nature of radio-selected BALS compared to other quasars}
The natural issue to consider given this sample is: How should this sample of radio-selected BAL quasars be treated in relation to other BAL quasars and radio-loud objects?  Because orientation can only be estimated in radio-loud BAL quasars at this time, the answer to this question has implications for the extension of such considerations to radio-quiet BAL quasars.

Our results clearly show that the rest-frame optical properties of radio-selected BAL quasars are, essentially, indistinguishable from radio-quiet BAL quasars. Visually, this is most striking from the comparison of composite spectra in Fig.~\ref{fig:composite}. \OIIIw\ emission is undetected, while optical \FeII\ emission is very strong.  This result is illustrated by Figure~\ref{fig:o3fe2rat} as well.  By corollary, the optical properties of radio-selected BAL quasars do not mimic those of radio-loud objects which tend to have strong \OIIIw\ and weak \FeII\ emission.  As noted earlier, \citet{dipompeo12b} also showed that the trends observed when comparing rest-frame spectral properties of BAL and non-BAL quasars persist whether the BALs are radio-loud or radio-quiet.  Though not conclusive, as a direct radio-loud/radio-quiet comparison is lacking, this does suggest that the findings here extend to other wavelengths.  

The radio luminosity, spectral index, and observed radio-loudness of our targets are summarized in Table 1. Under the canonical loud/quiet divisions (e.g., Kellermann et al. 1994, radio-loud: log$\,R^{*} >1$), all but two would be considered radio-loud, though none fall into the class of truly radio-powerful (log$\,R^{*} >2$) and large FR II sources, though such sources do exist \citep[e.g.,][]{brotherton02,gregg00,gregg06,miller09}. Though radio-selected, these BAL quasars tend to be considered radio-intermediate, rather than truly radio-loud. Nevertheless, the presence of radio emission is useful in gauging orientation. As already pointed out by \citet{becker00}, the radio properties of these BAL quasars are consistent with a range of jet orientations.  This finding has been confirmed for various other samples \citep{montenegro08,fine11}, as well as that of \citet{dipompeo11b,dipompeo12a} who find a range of but slight dependence on orientation (at the largest viewing angles one is more likely to see a BAL). Moreover the RL BAL sources presented in this work have a very high fraction of compact radio sources unresolved in FIRST (about 90\%) compared with the parent population they were found within (about 60\%), a fact difficult to reconcile without invoking parameters in addition to viewing angle (such as age).  

Should the results regarding the orientations of radio-selected BAL quasars be applied to all BAL quasars, the majority of which are truly radio-quiet? Our results from looking at optical properties suggest the answer is yes, as the radio-loud BAL quasars are more ``BAL-like'' than ``radio-loud-like''. While there have been suggestions in the past \citep[e.g.,][]{murray95,ogle99,elvis00} that BALs are seen in quasars at ``large'', more edge-on inclination angles, the evidence supporting this perspective is circumstantial, and contradictory to the radio properties of radio-selected BAL quasars \citep{shankar08, dipompeo11a,dipompeo12a}.  

Of note, hoewever, is that our sample of radio-selected BALs, like its parent sample from \citet{becker00}, has a higher fraction of  LoBALs and FeLoBALs than high-ionization BAL (HiBAL) quasars (see Table~\ref{tab:balprop}). There are no indications from the optical as to why these fractions would be different, either from the viewpoint of emission-line strengths, emission-line ratios, or continuum shape.  While optical polarization is generally higher in FeLo/LoBALs \citep[e.g.,][]{dipompeo10,dipompeo13a}, there appears to be no correlation with polarization and radio properties that would indicate FeLo/LoBALs are seen from the largest viewing angles.  Our findings here indicate that study of the full BAL subclass, from radio-quiet to radio-loud, should move away from the simplest orientation-based explanations.

Whether or not differences in geometry or age exist, differences in the radio properties of the BAL quasar population compared to other quasars clearly do exist given that at the largest viewing angles BALs are more likely to be observed \citep{dipompeo12a} and that the presence of BALs depends on radio power \citep{shankar08,dai12}.  Moreover the optical properties do not span the full range seen in non-BAL quasars, no matter the radio loudness, which we discuss in the next section.

\subsection{Eigenvector 1 versus accretion rate: BALQSOs and low \OIII/Fe$\,$\sc{ii}}
A picture is emerging in which BAL quasars, including radio-selected BAL quasars, display spectral properties at the well-defined, extreme end of EV1, with weak \OIII\ and strong \FeII\ \citep[e.g.,][]{bg92}.  In a sample selected to have weak \OIII, \citet{turnshek97} found a third of their objects had BALs, twice the normal 15\% \citep{gibson09}.  At the opposite extreme of EV1 spectral properties, BALs are simply not observed.  

From Figure~\ref{fig:hbfwhml}, the radio-selected BAL quasars in this sample have bolometric luminosities in between those of the \citet{yuan03} sample and the low redshift ($z<0.5$) objects from the Palomar-Green catalog \citep{bg92} and $z<0.89$ SDSS quasars from \citet{shen11}.  Moreover the estimates of Eddington ratio are not indicative of extreme accretion; only one object, Q$1031+3953$, has an Eddington ratio larger than unity.  This is curious since EV1 for low-redshift objects appears to be correlated strongly with Eddington ratio in the sense that objects with strong \FeII\ emission and weak \OIII\ emission should be accreting near the Eddington luminosity \citep{boroson02}.  Written before many RL BALs had been discovered, that work suggested that RL BALs should show extremely high Eddington ratios.  It is possible that EV1 behaves differently at different redshifts.  This would explain why the sample of \citet{boroson02}, with redshifts of a few tenths or less, shows a strong correlation between EV1 and Eddington ratio that is not confirmed in higher redshift samples, namely ours ($z\sim0.7-2.4$) and the \citet{yuan03} sample ($z\sim2$).

A clue to this mystery may be found in considering the dominant parameters that define EV1. In terms of \OIII\ and \FeII\ strength, our sample resembles radio-quiet BALs which would have small/low values of EV1. The \citet{boroson02} EV1 also involves other parameters, including the H$\beta$\ emission-line FWHM and formal radio-loudness. Our objects only represent an extreme of \OIII\ and the \FeII\ strengths, the dominant EV1 parameters, and not these other properties.

Multiple physical drivers for EV1 have been put forth in the literature.  At low-redshift, using the Palomar-Green sample with $z<0.5$, \citet{bg92} find that Eddington ratio dominates the variation of optical properties, and hence EV1, but at higher redshift this is not the case.  Since there is a common feature that \OIII\ emission is weak and \FeII\ emission is strong in BAL quasars of any redshift, \citet{yuan03} suggest that the availability of cold gas may be important in considering the relationship between BAL quasars and EV1. The anti-correlation between \OIII\ and \FeII\ emission may be a more fundamental relationship among quasars (i.e., at any redshift/luminosity). However, the entirety of EV1 as defined by low-z objects and its correlation with Eddington ratio might be secondary.  Further drivers include the suggestion that EV1 should be correlated with the ratio of optical to mid-infrared emission, which is indicative of the amount of reprocessed optical emission that should scale with the covering fraction of dust.  \citet{dipompeo13a} finds an infrared excess in radio-loud BAL quasars consistent with such a prediction.  \citet{wills99a} found that some of the line ratios associated with EV1 were diagnostic of gas density, and that property may be involved as well.  Our analysis, where we find that RL BAL quasars exhibit spectral properties on the extreme end of EV1 but not necessarily extremely high Eddington ratios, rules out the Eddington fraction as a driver of EV1, at least at high redshift.  With such varied evidence, it is clear that the physical driver of EV1 remains elusive.

\section{Conclusions}
\label{sec:conclusion}

We have observed the rest-frame optical properties of eight radio-selected BAL quasars from \citet{becker00} using the NASA IRTF/SpeX instrument. Using the {\sc specfit} package, we have separated the power-law continuum, \FeII\ emission, \OIII\ emission, H$\alpha$\ and H$\beta$\ emission.  Furthermore, we have estimated the black hole masses and Eddington ratios of these quasars.  Radio-selected BAL quasars have similar optical properties (weak \OIII, strong \FeII) as radio-quiet BAL quasars and are dissimilar to RL non-BAL quasars in this sense.  In short, RL BAL quasars are more ``BAL-like'' than ``RL-like''.  As these objects are not extreme accretors, we conclude that the properties defining EV1 at low redshift may not be directly extrapolated to high redshift/luminosity objects. The anti-correlation between \OIII\ and \FeII\ emission may be fundamental with BAL quasars lying at one extreme, but this is not necessarily a result of Eddington ratio.

\section*{Acknowledgments}

We wish to thank Bev Wills for helpful discussions and assistance with reduction of the IRTF data and the anonymous referee for helpful and detailed comments that improved the presentation of this work. This work was part of the Wyoming SURAP REU program funded by the National Science Foundation grant ASTR-0353760. We acknowledge support from the US National Science Foundation through grant AST 05-07781. Z. S. acknowledges the support from the National Natural Science Foundation of China through grant 10633040.

This publication makes use of data products from the Two Micron All Sky Survey, which is a joint project of the University of Massachusetts and the Infrared Processing and Analysis Center/California Institute of Technology, funded by the National Aeronautics and Space Administration and the National Science Foundation.

Funding for the SDSS and SDSS-II has been provided by the Alfred P. Sloan Foundation, the Participating Institutions, the National Science Foundation, the U.S. Department of Energy, the National Aeronautics and Space Administration, the Japanese Monbukagakusho, the Max Planck Society, and the Higher Education Funding Council for England. The SDSS Web Site is http://www.sdss.org/.

The SDSS is managed by the Astrophysical Research Consortium for the Participating Institutions. The Participating Institutions are the American Museum of Natural History, Astrophysical Institute Potsdam, University of Basel, University of Cambridge, Case Western Reserve University, University of Chicago, Drexel University, Fermilab, the Institute for Advanced Study, the Japan Participation Group, Johns Hopkins University, the Joint Institute for Nuclear Astrophysics, the Kavli Institute for Particle Astrophysics and Cosmology, the Korean Scientist Group, the Chinese Academy of Sciences (LAMOST), Los Alamos National Laboratory, the Max-Planck-Institute for Astronomy (MPIA), the Max-Planck-Institute for Astrophysics (MPA), New Mexico State University, Ohio State University, University of Pittsburgh, University of Portsmouth, Princeton University, the United States Naval Observatory, and the University of Washington

\bibliographystyle{/Users/jrunnoe/Library/texmf/bibtex/bst/mn2e}
\bibliography{all.041513}
\clearpage

\label{lastpage}
\end{document}